\documentclass[a4paper,11pt]{article}
\usepackage{pos}

\newcommand{\be}{\begin{equation}}
\newcommand{\ee}{\end{equation}}
\newcommand{\bea}{\begin{eqnarray}}
\newcommand{\eea}{\end{eqnarray}}
\newcommand{\MSbar}{{\overline{\rm MS}}}
\usepackage{IEEEtrantools}
\usepackage{subcaption}

\title{Non-perturbative renormalization of quark and gluon operators using a gauge-invariant scheme}
\ShortTitle{Non-perturbative renormalization of quark and gluon operators using GIRS}

\author*[a]{G.~Spanoudes}
\author[a,b]{C.~Alexandrou}
\author[a]{J.~Finkenrath}
\author[a,b]{K.~Hadjiyiannakou}
\author[b]{H.~Panagopoulos}
\author[a]{S.~Yamamoto}

\affiliation[a]{Computation-based Science and Technology Research Center, The Cyprus Institute, 20 Kavafi Str., Nicosia 2121, Cyprus}
\affiliation[b]{Department of Physics, University of Cyprus, Nicosia, CY-1678, Cyprus}

\emailAdd{g.spanoudis@cyi.ac.cy}

\abstract{We present preliminary results for the renormalization functions (RFs) of a number of quark and gluon operators studied in lattice QCD using a gauge-invariant renormalization scheme (GIRS). GIRS is a variant of the coordinate-space renormalization prescription, in which Green's functions of gauge-invariant operators are calculated in position space. A novel aspect is that summations over different time slices of the  positions of the operators are employed in order to reduce the statistical noise in lattice simulations. We test the reliability of this scheme by calculating RFs for the vector one-derivative quark bilinear operator, which enters the average momentum fraction of the nucleon. We use $N_f=4$ degenerate twisted mass clover-improved fermion ensembles of different volumes and lattice spacings. We also present first results of applying GIRS when operator mixing occurs: the mixing coefficients of the gluon and quark singlet energy-momentum tensor operators are evaluated by imposing appropriate renormalization conditions on the lattice.
\begin{center}
\includegraphics[scale=0.45]{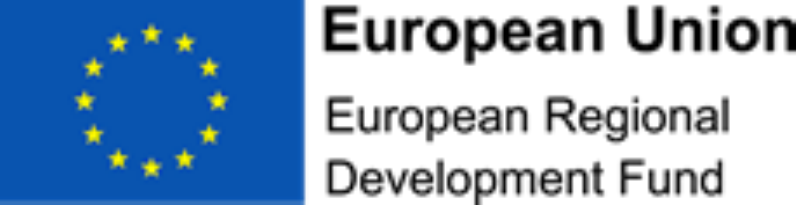}
\includegraphics[scale=0.45]{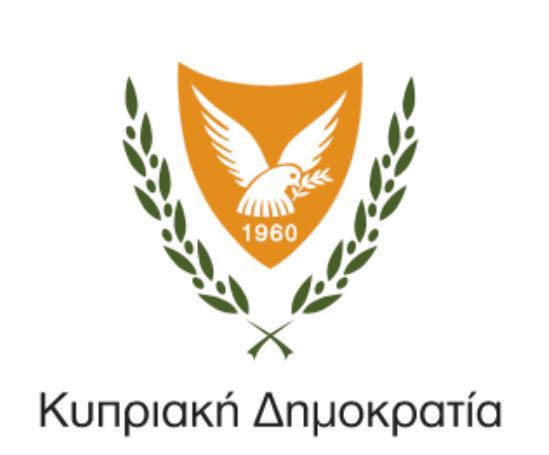}
\includegraphics[scale=0.45]{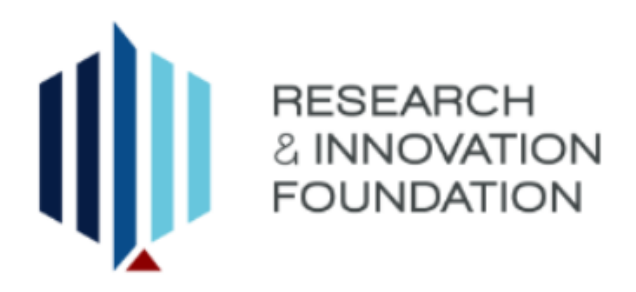}
\includegraphics[scale=0.45]{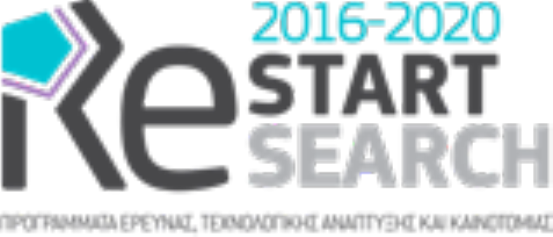}
\end{center}}

\FullConference{%
  The 39th International Symposium on Lattice Field Theory (Lattice2022),\\
  8-13 August, 2022, \\
  Rheinische Friedrich-Wilhelms-Universität Bonn, \\
  Bonn, Germany 
}

\begin{document}
\maketitle

\section{Introduction}
\label{secI}

Renormalization of quantum-field operators is an essential ingredient when studying physical quantities in Hadron Physics, which are not conserved currents. It is the procedure that we use to remove ultra-violet (UV) divergences from correlation functions in order to make contact to physical observables. There are various ways of renormalizing an operator, called schemes, by imposing appropriate conditions on particular Green's functions. In phenomenology, the typical renormalization scheme used in the analysis of experimental data is the modified minimal subtraction ($\MSbar$) scheme. $\MSbar$ is originally defined in dimensional regularization at each order of perturbation theory; it is a minimal scheme in the sense of removing only pole terms\footnote{In addition, finite terms proportional to $\ln(4 \pi) - \gamma_E$, where $\gamma_E$ is Euler's constant, are also removed in order to have smaller coefficients in the perurbative expansion.} appearing in correlation functions when the space-time dimension takes its physical value, $d=4$. Such a minimal scheme is difficult to implement in a non-perturbative way. Thus, a direct non-perturbative evaluation of $\MSbar$-renormalized correlation functions on the lattice (via Monte Carlo simulations) is not feasible.

An indirect way of obtaining non-perturbative results in $\MSbar$ on the lattice has been successfully implemented in the last two decades through the use of an intermediate scheme, which is applicable in both lattice and continuum regularizations. The main idea is to calculate non-perturbative renormalization functions (RFs) in the intermediate scheme on the lattice and then to convert to $\MSbar$ by using perturbative regularization-independent matching factors; the latter can be calculated in dimensional regularization at high perturbative order. Some popular intermediate schemes extensively employed in lattice-QCD simulations are the RI$'$/MOM and the Schr{\" o}dinger functional schemes. In this work, we investigate an alternative intermediate scheme in which only gauge-invariant Green's functions are considered; we refer to it as gauge-invariant renormalization scheme (GIRS), and it is an extension of older coordinate-space prescriptions~\cite{Gimenez:2004me, Chetyrkin:2010dx, Cichy:2012is, Tomii:2018zix}. Our study focuses on the application of this scheme in both multiplicative renormalization and renormalization in the presence of operator mixing; in the latter case, GIRS is more suitable than other schemes, since it simplifies the mixing pattern (see Sec.~\ref{secIII}). 

In GIRS, conditions are imposed on Green's functions of products of gauge-invariant operators $\mathcal{O}$ at different space-time points (in a way as to avoid potential contact singularities) and in the chiral limit. For instance, a typical condition in the case of multiplicatively renormalizable operators has the following form (see, e.g., Ref.~\cite{Gimenez:2004me}):
\begin{equation}
{(Z_{\mathcal{O}}^{\rm GIRS})}^2 \langle \mathcal{O} (x) \mathcal{O} (y) \rangle|_{x-y = z} = \langle \mathcal{O} (x) \mathcal{O} (y) \rangle^{\rm tree}|_{x-y = z},
\label{GIRScondition}
\end{equation}
where $z$ is a non-zero renormalization 4-vector scale within the renormalization window: $a \ll |z| \ll \Lambda_{\rm QCD}^{-1}$ (where $a$ is the lattice spacing and $\Lambda_{\rm QCD}$ is the QCD physical scale). The window must be wide enough in order to keep lattice artifacts under control and at the same time to ensure reliability of continuum perturbation theory. When operator mixing occurs, we need to consider a set of conditions involving more than one Green's functions of two or more gauge-invariant operators, each of which has a similar form to Eq. (\ref{GIRScondition}), i.e., the renormalized Green's functions are set to their tree-level values when the operators' space-time separations equal to specific reference scales (see Sec.~\ref{secIII}).

The main advantages of GIRS compared to other schemes are: \\
{\bf 1.} Due to the gauge-independent nature of GIRS, gauge fixing is not needed. \\
{\bf 2.} Gauge-non-invariant (GNI) operators have vanishing correlation functions in GIRS; thus, when mixing occurs, all GNI operators [Becchi-Rouet-Stora-Tyutin (BRST) variations and operators which vanish by the equations of motion], which can mix with gauge-invariant operators, can be safely excluded from the renormalization procedure, leading to a reduced set of mixing operators. \\
{\bf 3.} Contact terms are automatically excluded by the definition of the GIRS correlation functions. \\
{\bf 4.} Perturbative matching of GIRS and $\MSbar$ scheme is possible at high perturbative order (in most cases).

A particular aspect of the present study is that summations over time slices of the operator insertion points are employed in order to reduce the statistical noise in lattice simulations:
\begin{equation}
  {(Z_{\mathcal{O}}^{\rm GIRS})}^2 \ \sum_{\vec{x},\vec{y}} \ \langle \mathcal{O} (\vec{x},x_4) \mathcal{O} (\vec{y},y_4) \rangle \Big|_{x_4 - y_4 = t} = \sum_{\vec{x},\vec{y}} \ \langle \mathcal{O} (\vec{x},x_4) \mathcal{O} (\vec{y},y_4) \rangle^{\rm tree} \Big|_{x_4 - y_4 = t},
 \label{cond}
 \end{equation}
where $t \neq 0$ is the GIRS scale.

In what follows, we consider two applications of GIRS, one in the multiplicative renormalization of the vector one-derivative quark bilinear operator (Sec.~\ref{secII}) and the other in the study of mixing of the QCD traceless gluon and quark flavor-singlet energy-momentum tensor (EMT) operators (Sec.~\ref{secIII}). The hadron matrix elements of these operators enter the calculations of the average momentum fraction and the spin contribution of the constituent gluons and quarks inside the hadron. We use several $N_f=4$ twisted mass and clover-improved fermion ensembles of different volumes and lattice spacings. The details are given in Table~\ref{tab:ensembles}.
 
\section{Application of GIRS to the renormalization of the non-singlet vector derivative quark bilinear operator}
\label{secII}

First, we test the reliability of GIRS by calculating the RFs for the non-singlet vector derivative operator:
\begin{equation}
  \mathcal{O}_{{\rm DV}_{\mu \nu}} (x) = \bar{d} (x) \gamma_{\{\mu} \overleftrightarrow{D}_{\nu \}} u (x),
\end{equation}
where $\overleftrightarrow{D}_\mu$ is the symmetrized covariant derivative and $\{ \ldots \}$ denotes the symmetrization over Lorentz indices $\mu, \nu$ and subtraction of the trace. The operator is chosen to be traceless in order to avoid mixing with lower dimensional operators. On the lattice, where Lorentz symmetry is replaced by hypercubic symmetry, diagonal ($\mu = \nu$) and non-diagonal ($\mu \neq \nu$) components of traceless symmetric operators belong to different representations of the hypercubic group, and thus, they renormalize differently. In this study, we focus on the renormalization of the non-diagonal (nd) components.

The strategy that we follow for extracting the RFs of the non-diagonal vector derivative operator using GIRS is summarized below: \\
{\bf 1.} We calculate the following Green's function of two vector derivative operators at different time slices:
\begin{equation}
  G \left(\frac{t_s}{a} - \frac{t_0}{a}\right) \equiv \frac{1}{6} \sum_{\vec{x},\vec{y}} \sum_{i \neq j} \left\langle \mathcal{O}_{{\rm DV}_{i j}} \left(\vec{x},\frac{t_s}{a}\right) \ \mathcal{O}^{\dagger}_{{\rm DV}_{i j}} \left(\vec{y},\frac{t_0}{a}\right) \right\rangle,
  \label{bareGF}
\end{equation}
for different values of the source-sink time separations $(t_s - t_0)/a$. To alleviate the statistical noise, we take averages over different Lorentz components of the operators due to the hypercubic symmetry. Note that in the averages we have not included temporal components, since they give vanishing contributions in the continuum, where the summations over the spatial components of the operators’ positions are replaced by integrations. \\
{\bf 2.} In order to reduce the systematics, we eliminate discretization errors  from the Green’s function by removing tree-level artifacts calculated in lattice perturbation theory. \\
{\bf 3.} We impose the following renormalization condition and we extract the RF ($Z_{{\rm DV}_{\rm nd}}^{\rm GIRS}$) of the non-diagonal vector derivative operator in GIRS:
\begin{equation}
  {(Z_{{\rm DV}_{\rm nd}}^{\rm GIRS})}^2 \times G \left(\frac{t_s}{a} - \frac{t_0}{a}\right)|_{t_s - t_0 = t} = G^{\rm tree} \left(\frac{t_s}{a} - \frac{t_0}{a}\right) |_{t_s - t_0 = t},
\end{equation}
where $G^{\rm tree} \left(t/a\right) = (3 N_c)/(40 \pi^2 {|t/a|}^5)$, $N_c$ is the number of colors. \\
{\bf 4.} We convert to the $\overline{\rm MS}$ scheme at the reference scale of 2 GeV by calculating conversion factors $C_{{\rm DV}_{\rm nd}}^{{\rm GIRS} \to \overline{\rm MS}}$ and evolution functions $R_{\rm DV}^{\overline{\rm MS}} (\mu_1, \mu_2)$ in the continuum at some perturbative order:
\begin{equation}
  C_{{\rm DV}_{\rm nd}}^{{\rm GIRS} \to \overline{\rm MS}} \equiv \frac{Z_{{\rm DV}_{\rm nd}}^{\overline{\rm MS}}}{Z_{{\rm DV}_{\rm nd}}^{\rm GIRS}}, \qquad R_{\rm DV}^{\overline{\rm MS}} (\mu_1, \mu_2) = \exp \left( - \int_{g^{\overline{\rm MS}} (\mu_1)}^{g^{\overline{\rm MS}} (\mu_2)} dg \frac{\gamma^{\overline{\rm MS}}_{\mathcal{O}_{\rm DV}} (g)}{\beta^{\overline{\rm MS}} (g)} \right),
\end{equation}
where $\beta^{\overline{\rm MS}} (g)$, $\gamma^{\overline{\rm MS}}_{\mathcal{O}_{\rm DV}} (g)$ are the beta function and the anomalous dimension of the operator, respectively. \\
{\bf 5.} We take a plateau fit over the GIRS scale $t$, since the $\overline{\rm MS}$ renormalization factor must be independent on the initial scale.
  
  In our analysis, we employ $N_f = 4$ dynamical ensembles, generated by the Extented Twisted Mass Collaboration (ETMC), at three different lattice spacings (a) and different volumes ($L^3 \times T$), using four degenerate light maximally twisted-mass/clover quarks and Iwasaki-improved gluons. The goal of our study is to extract RFs for different values of the lattice spacing in order to be able to perform a continuum extrapolation on the nucleon matrix elements. The parameters of the ensembles are given in Table~\ref{tab:ensembles}. The values of the lattice spacing have been determined using the nucleon mass.
  \begin{table}[!h]
  \centering
\begin{tabular} {l | l | l | l | l | l | l}
  \hline
  \hline
   Ensemble & $\beta$  & $a$ (fm) & $L^3 \times T$ & $a \mu$ & $\kappa$ & $c_{\rm SW}$ \\
   \hline
   cB4.060.12  & 1.778 & 0.080 & $12^3 \times 24$ & 0.006 & 0.1393050 & 1.6900 \\
   cB4.060.16  & 1.778 & 0.080 & $16^3 \times 32$ & 0.006 & 0.1393050 & 1.6900 \\
   cB4.060.24  & 1.778 & 0.080 & $24^3 \times 48$ & 0.006 & 0.1393050 & 1.6900 \\
   cC4.050.24  & 1.836 & 0.069 & $24^3 \times 48$ & 0.005 & 0.1386735 & 1.6452 \\
   cC4.050.32  & 1.836 & 0.069 & $32^3 \times 64$ & 0.005 & 0.1386735 & 1.6452 \\
   cD4.040.48  & 1.900 & 0.058 & $48^3 \times 96$ & 0.004 & 0.13793128 & 1.6112 \\ 
   \hline
\end{tabular}
\caption{Ensembles and their parameters used for the calculations: $\beta = (2 N_c/g^2)$, where $g$ is the running coupling, lattice spacing (a), lattice volume ($L^3 \times T$), twisted-mass parameter $(a \mu)$, hopping parameter $\kappa$ and clover coefficient $c_{\rm SW}$.}
    \label{tab:ensembles}
    \end{table}
The number of configurations in each ensemble varies between 100 to 200.

\begin{figure}
    \hspace*{-0.5cm}\includegraphics[scale=0.54]{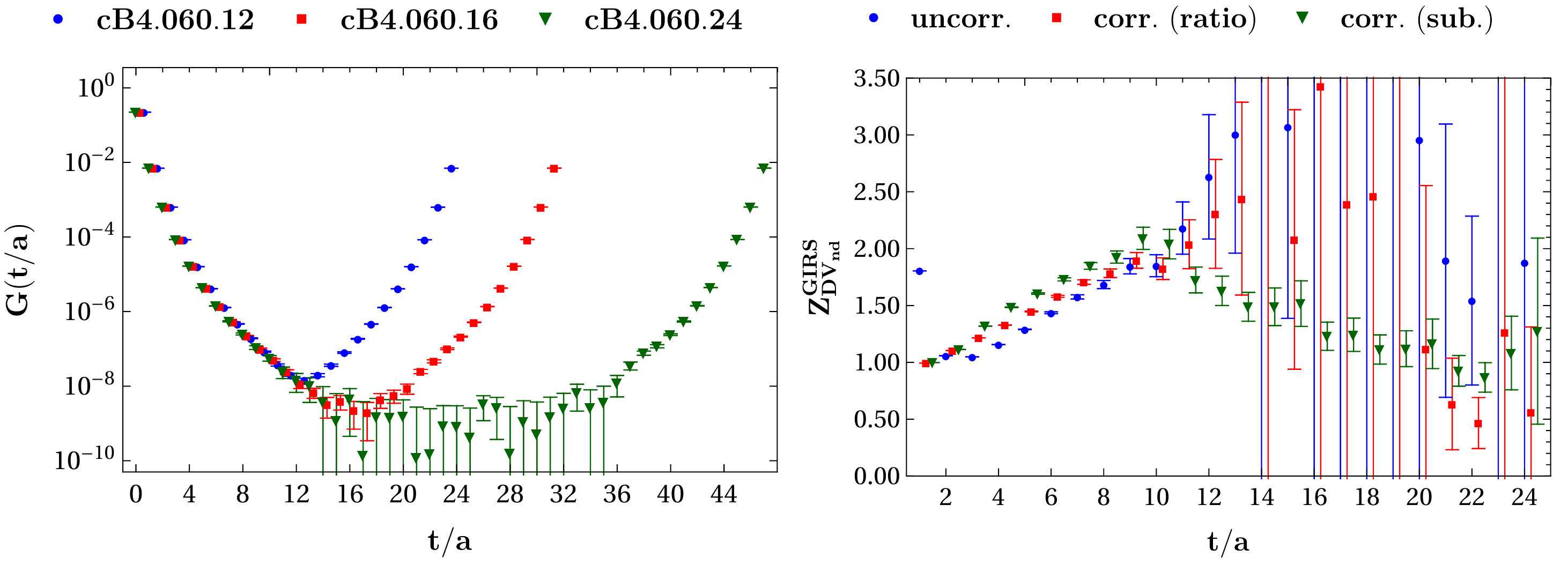}
    \caption{{\bf Left:} Plot of $G (t/a)$ as a function of $t/a$ for different lattice sizes (ensembles: cB4.060.12, cB4.060.16, cB4.060.24).
{\bf Right:} Plot of $Z_{{\rm DV}_{\rm nd}}^{\rm GIRS}$ as a function of $t/a$ for the ensemble cB4.060.24 with and without tree-level corrections by employing ratio and subtraction methods explained in the text.}
    \label{Fig:GFFvst}
\end{figure}

To exemplify the procedure we provide step-by-step results for the ensemble cB4.060.24. In the first step, the bare Green’s function of Eq. \eqref{bareGF} is calculated using the stochastic method of one-end trick, described in Ref.~\cite{McNeile:2006bz}. The number of stochastic sources employed in each ensemble varies between 5 to 12. Time dilution is also performed to further improve the estimator. Periodic boundary conditions have been implemented for the gluon fields and (anti-)periodic boundary conditions in the (temporal) spatial direction for the quark fields. In order to check for possible volume effects, we compare the calculated Green's function for different lattice sizes (ensembles: cB4.060.12, cB4.060.16, cB4.060.24). From the left plot of Fig.~\ref{Fig:GFFvst}, we conclude that there are no significant volume effects especially for the smallest values of the source-sink time separation $t/a$. In what follows, we focus on the larger volume ensemble cB4.060.24. To further improve the statistical signal we perform the average: $[G \left(t/a\right) + G \left(T/a -t/a)\right)]/2$, where $T$ is the time component of the lattice volume, since the Green's function is even under time reversal due to the use of periodic/anti-periodic boundary conditions.

In the second step, we apply two different methods in order to reduce tree-level discretization errors: \\
{\bf 1. Ratio method:} $G^{\rm corr. \ (tree \ ratio)} (t/a) = G(t/a) \times R(t/a)$, \\
{\bf 2. Subtraction method:} $G^{\rm corr. \ (tree \ subtracted)} (t/a) = G(t/a) - D(t/a)$, \\
where $R(t/a) \equiv [G^{\rm tree, \ cont.}(t/a)]/[G^{\rm tree, \ lat.}(t/a)]$, $D(t/a) \equiv G^{\rm tree, \ lat.}(t/a) - G^{\rm tree, \ cont.}(t/a)$, and $G^{\rm tree, \ cont.}(t/a) (G^{\rm tree, \ lat.}(t/a))$ is the continuum (lattice) tree-level value of $G (t/a)$; the perturbative calculation on the lattice has been performed to all orders in the lattice spacing for the finite volume $24^3 \times 48$. The two methods differ by higher-loop discretization errors. We test both methods by extracting the GIRS RF $[Z_{{\rm DV}_{\rm nd}}^{\rm GIRS}]$. In the right plot of Fig.~\ref{Fig:GFFvst}, we observe that both correction methods work for small values of $t/a$, while for larger values only the subtraction method gives reliable results. However, the discrepancy between the two methods indicate that higher-loop discretization errors are still important. For the next steps we consider both methods, since the renormalization window lies in small values of $t/a$.

\begin{figure}
\centering
  \includegraphics[scale=0.7]{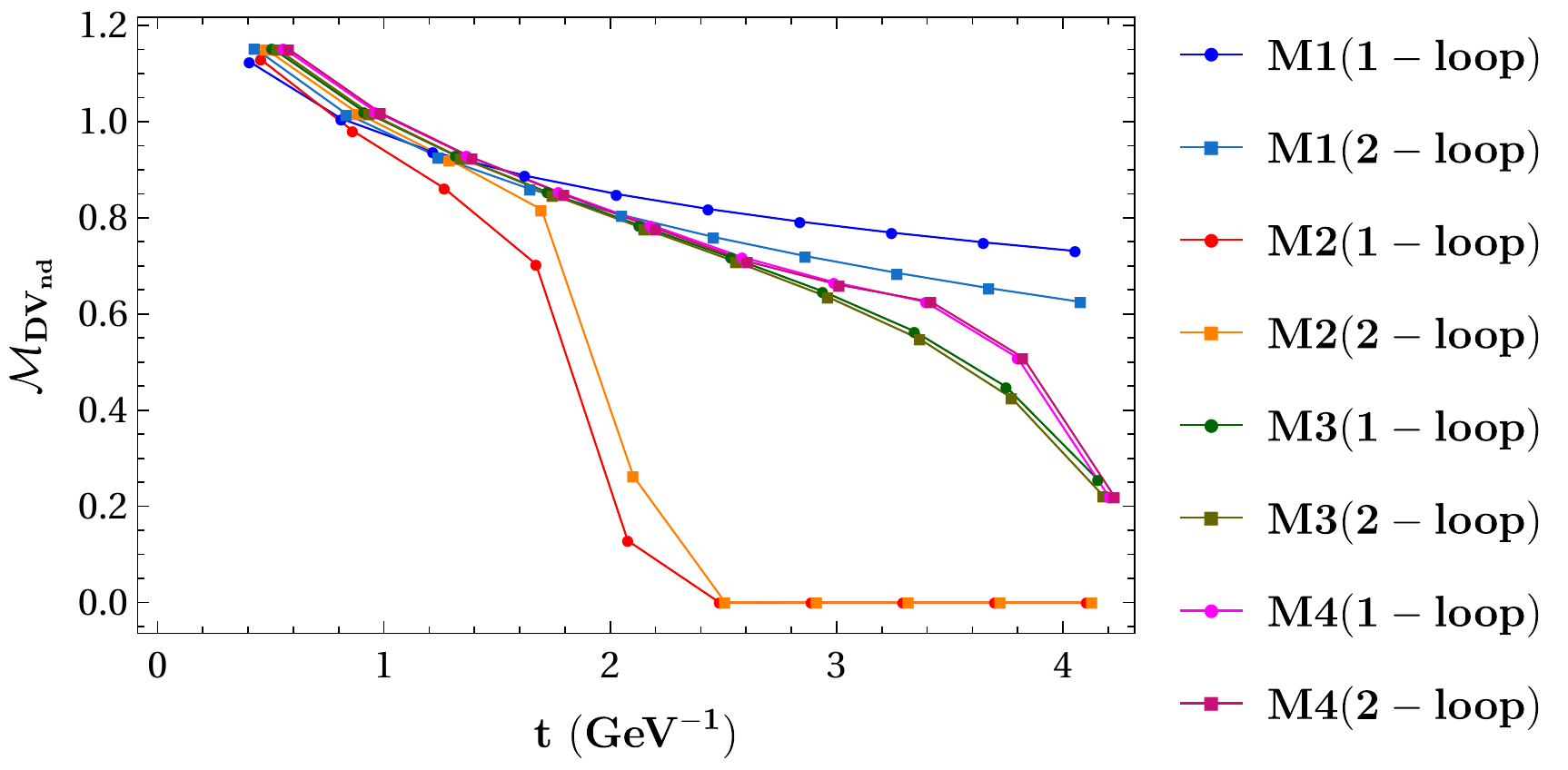}
  \caption{Plot of the matching factor between GIRS and $\overline{\rm MS}$ schemes at the reference scale of 2 GeV, by applying 4 different methods (M1 - M4) descibed in the text.}
  \label{fig:CGIRSMSbar}
\end{figure}

In the next step, we perform the conversion to the $\overline{\rm MS}$ scheme at the reference scale of 2 GeV, by using 4 different methods, which differ by higher-loop corrections: \\
{\bf 1. M1:} We employ the two-loop conversion factor, $C_{{\rm DV}_{\rm nd}}^{{\rm GIRS} \to \overline{\rm MS}}$, that we have calculated in continuum perturbation theory\footnote{The one-loop result is given in Ref.~\cite{Costa:2021iyv}, while the two-loop result will be provided in a forthcoming publication.} for an arbitrary renormalization scale $\bar{\mu}$, and we set the scale directly to 2 GeV. \\
{\bf 2. M2:} We convert to $\overline{\rm MS}$ at an intermediate scale of $\bar{\mu} = 1/t$ in order to eliminate the logarithmic terms $\ln (\bar{\mu}^2 t^2)$ appearing in the conversion factor (see~\cite{Costa:2021iyv}), and then we make use of the 3-loop expression for the anomalous dimension of the vector derivative operator, calculated in Refs.~\cite{baikov2015massless,Herzog:2018kwj}, in order to evolve to 2 GeV. \\
{\bf 3. M3:} Method 3 is similar to method 2 with the difference of tuning the intermediate scale $\bar{\mu}$ in order to take more convergent conversion factors. This can be achieved by comparing the size of the one-loop and two-loop contributions to the conversion factor, when varying the intermediate scale. We found that the optimal value of the intermediate scale is $\bar{\mu} \approx 2.18/t$. \\
{\bf 4. M4:} In method 4, we generalize method 3 by adding a second scale through the evolution of the running coupling ($g^{\overline{\rm MS}}$) to a different scale (as first proposed in Ref.~\cite{Tomii:2016xiv}). In this way, we need to tune two different variables in order to get even better convergence of the conversion factor. The optimal value for the second scale is found to be $\sim 1.5/t$. \\
All methods are tested by comparing the one-loop and two-loop contributions to the overall matching factor (conversion factor and/or evolution function), $\mathcal{M}_{{\rm DV}_{\rm nd}}$, as shown in Fig.~\ref{fig:CGIRSMSbar}. As we observed, M3 and M4 give more compatible results between the one and two-loop matching factors, and thus, we proceed only with these two methods.

\begin{figure}
\centering
    \includegraphics[scale=0.75]{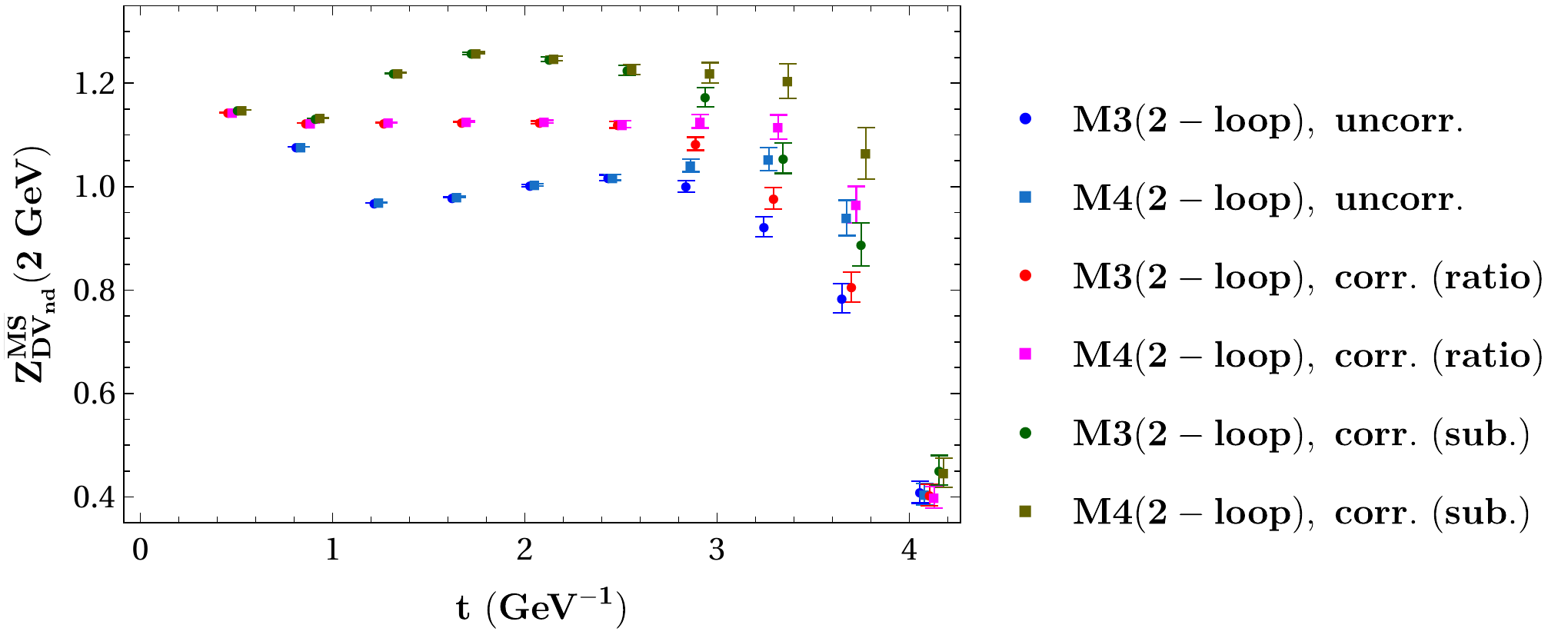}
  \caption{Plot of $Z_{{\rm DV}_{\rm nd}}^{\overline{\rm MS}} (2 \ {\rm GeV})$ as a function of $t/a$ by employing methods M3 and M4 of the matching for both uncorrected and corrected (using ratio and subtraction methods) renormalization factor.}
  \label{ZMSbar_B}
\end{figure}

\begin{figure}
\centering
  \includegraphics[scale=0.78]{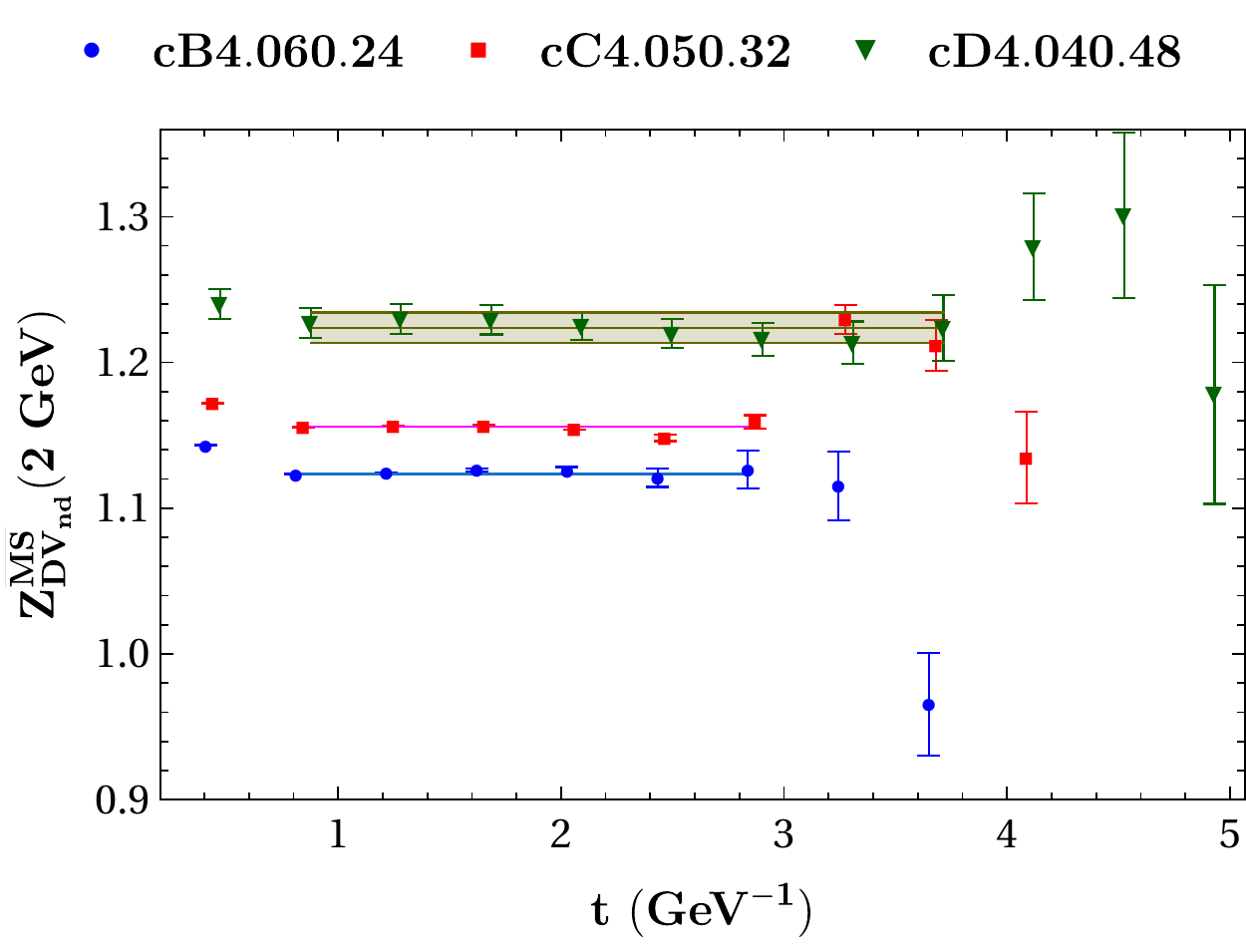}
  \caption{Plateau fit of $Z_{{\rm DV}_{\rm nd}}^{\overline{\rm MS}} (2 \ {\rm GeV})$ for the ensembles cB4.060.24, cC4.050.32 and cD4.040.48.}
  \label{ZMSbar_B_C_D}
  \end{figure}

In Fig.~\ref{ZMSbar_B} we show results for the RF in the $\overline{\rm MS}$ scheme by employing the two methods of the matching for both uncorrected and corrected renormalization factor using the ratio and the subtraction method for reducing discretization errors. As we see, a clearer and a larger plateau can be obtained from the ratio method and method 4 of the matching. By employing these methods, we take the plateau fit as the final step (see Fig.~\ref{ZMSbar_B_C_D}). The interval used for the fit is (0.5 - 2.9) ${\rm GeV}^{-1}$, which lies in the renormalization window.

Omiting intermediate results, we also present, in Fig.~\ref{ZMSbar_B_C_D}, the final step of taking the plateau fit for the remaining ensembles C and D. In Table~\ref{tab:results}, we provide the final values of the renormalization factors after the plateau fit. Systematic errors are estimated by varying the intermediate scales
 \begin{table}[!h]
  \centering
  \begin{tabular} {l | l | l}
    \hline
    \hline
   Ensemble & GIRS & RI/MOM \\
   \hline
cB4.060.24 & 1.1234(1)(46) & 1.1209(5)(142) \\
cC4.050.32 & 1.1558(1)(43) & 1.1490(3)(111) \\
cD4.040.48 & 1.2239(104)(131) & 1.1995(4)(113) \\
   \hline
  \end{tabular}
  \caption{Comparison of $Z_{{\rm DV}_{\rm nd}}^{\overline{\rm MS}} (2 \ {\rm GeV})$ coming from the GIRS and RI$'$/MOM prescriptions. The error in the first (second) parenthesis is statistical (systematic).}
  \label{tab:results}
 \end{table}
in the matching factor and by varying the plateau fit interval. We compare the results with those coming from a similar study using the RI$'$/MOM scheme for the same ensembles. The results of the two schemes are compatible (within error) for all ensembles.

\section{Application of GIRS to the renormalization of the QCD traceless energy-momentum tensor}
\label{secIII}

As a follow-up, we proceed with the application of GIRS to the renormalization of the QCD traceless gluon and quark energy-momentum tensor (EMT) operators:
\begin{equation}
    \overline{T}^{g}_{\mu \nu} (x) = -2 {\rm Tr} \ [F_{\rho \{\mu} (x) \ F_{\nu \} \rho} (x)], \hspace{1.3cm} \overline{T}^{q}_{\mu \nu} (x) = \sum_{f=1}^{N_f} \bar{\psi}_f (x) \gamma_{\{\mu} \overleftrightarrow{D}_{\nu \}} \psi_f (x),
\end{equation}
The calculation of the renormalized hadron matrix elements of these operators are very important for extracting useful key observables. In particular, by decomposing the nucleon matrix elements of the EMT operators into generalized form factors~\cite{Ji:1998pc} and taking the forward limit, we can extract the gluon and quark average momentum fractions $\langle x \rangle^{g(q)}$, as well as the gluon and quark contributions to the nucleon spin $J^{g(q)}$:
\begin{eqnarray}
   &&\hspace{-0.6cm}\langle N(\vec{p'},s') | {\overline{T}}^{g(q)}_{\mu \nu} | N(\vec{p},s) \rangle = \nonumber \\
 && \hspace{-0.4cm}{\overline{u}}_N (\vec{p'},s') \left[ A_{20}^{g(q)} (q^2) \ i \gamma_{\{ \mu} P_{\nu \}} + B_{20}^{g(q)} (q^2) \frac{i P_{\{ \mu} \sigma_{\nu \} \rho} q_\rho}{2 m_N} + C_{20}^{g(q)} (q^2) \frac{q_{\{ \mu} q_{\nu \}}}{m_N} \right] u_N (\vec{p},s), \qquad
 \end{eqnarray}
where $P \equiv (p' + p) / 2$, $\ q \equiv p' - p$, and $\langle x \rangle^{g(q)} = A_{20}^{g(q)} (0)$, $\ J^{g(q)} = [A_{20}^{g(q)} (0) +  B_{20}^{g(q)} (0)]/2$. The ultimate goal of our study is to confirm the momentum and spin sum rules: $\langle x \rangle^{g} + \langle x \rangle^{q} = 1$, $J^g + J^q = 1/2$.

A difficulty in studying the renormalization of these operators is that mixing is present; the two operators along with three GNI operators~\cite{Caracciolo:1991cp} mix among themselves, as they have the same transformations under Euclidean rotational (or hypercubic, on the lattice) symmetry. However, by using the gauge-invariant scheme the mixing problem is simplified, since the GNI operators are automatically excluded from the renormalization
procedure. Then, for the renormalization of the two remaining operators, we need to construct a $2 \times 2$ mixing matrix:
\begin{equation}
 \begin{pmatrix}
\bar{T}^{g,R}_{\mu \nu} \\
\bar{T}^{q,R}_{\mu \nu}
 \end{pmatrix} =
 \begin{pmatrix}
Z_{gg} & Z_{gq} \\
Z_{qg} & Z_{qq}
 \end{pmatrix}
 \begin{pmatrix}
\bar{T}^g_{\mu \nu} \\
\bar{T}^q_{\mu \nu}
 \end{pmatrix}
\end{equation}

The calculation of all mixing matrix elements requires a total of four conditions involving correlation functions of $\overline{T}^{g(q)}_{\mu \nu}$. Three conditions can be obtained by considering two-point functions between the two EMT operators. A fourth condition can be obtained by considering three-point functions among an EMT operator and two lower dimensional operators, e.g., two fermion bilinears\footnote{Two-point functions between an EMT operator and one fermion bilinear operator vanish due to trace algebra or charge conjugation symmetry.}. Below, we propose a solvable system of conditions for the case of non-diagonal ($\mu \neq \nu$) EMT operators (in terms of renormalized operators):
\begin{eqnarray}
  &&\frac{1}{6} \sum_{\vec{x},\vec{y}} \sum_{i \neq j} \left\langle {\overline{T}^g_{ij}}^{\rm GIRS} (\vec{x},\frac{t_s}{a}) \ {\overline{T}^g_{ij}}^{\rm GIRS} (\vec{y},\frac{t_0}{a}) \right\rangle|_{t_s - t_0 = t} = \frac{3 (N_c^2 - 1)}{20 \pi^2 |t/a|^5}, \\
  &&\frac{1}{6} \sum_{\vec{x},\vec{y}} \sum_{i \neq j} \left\langle {\overline{T}^q_{ij}}^{\rm GIRS} (\vec{x},\frac{t_s}{a}) \ {\overline{T}^q_{ij}}^{\rm GIRS} (\vec{y},\frac{t_0}{a}) \right\rangle|_{t_s - t_0 = t} = \frac{3 N_c N_f}{40 \pi^2 |t/a|^5}, \\
  &&\frac{1}{6} \sum_{\vec{x},\vec{y}} \sum_{i \neq j} \left\langle {\overline{T}^g_{ij}}^{\rm GIRS} (\vec{x},\frac{t_s}{a}) \ {\overline{T}^q_{ij}}^{\rm GIRS} (\vec{y},\frac{t_0}{a}) \right\rangle|_{t_s - t_0 = t} = 0, \\
  &&\frac{1}{6} \sum_{\vec{x},\vec{y},\vec{z}} \sum_{i \neq j} \left\langle \mathcal{O}_{\gamma_i}^{\rm GIRS} (\vec{x},\frac{t_s}{a}) {\overline{T}^g_{ij}}^{\rm GIRS} (\vec{y},\frac{t_i}{a}) \ {\mathcal{O}_{\gamma_j}^{\rm GIRS}}^{\dagger} (\vec{z},\frac{t_0}{a}) \right\rangle|_{\begin{smallmatrix}
      t_s - t_0 = 2t, \\
      t_i - t_0 = t
  \end{smallmatrix}} = 0,
\end{eqnarray}
where $\mathcal{O}_{\gamma_i} = \bar{d} \gamma_i u$ is the flavor non-singlet vector operator. As in the non-singlet case, we only consider spatial Lorentz components of the operators in order to avoid vanishing contributions in the continuum, which can lead to a non-solvable sytem of equations.

\begin{figure}
  \centering
  \hspace*{-0.5cm}\includegraphics[scale=0.54]{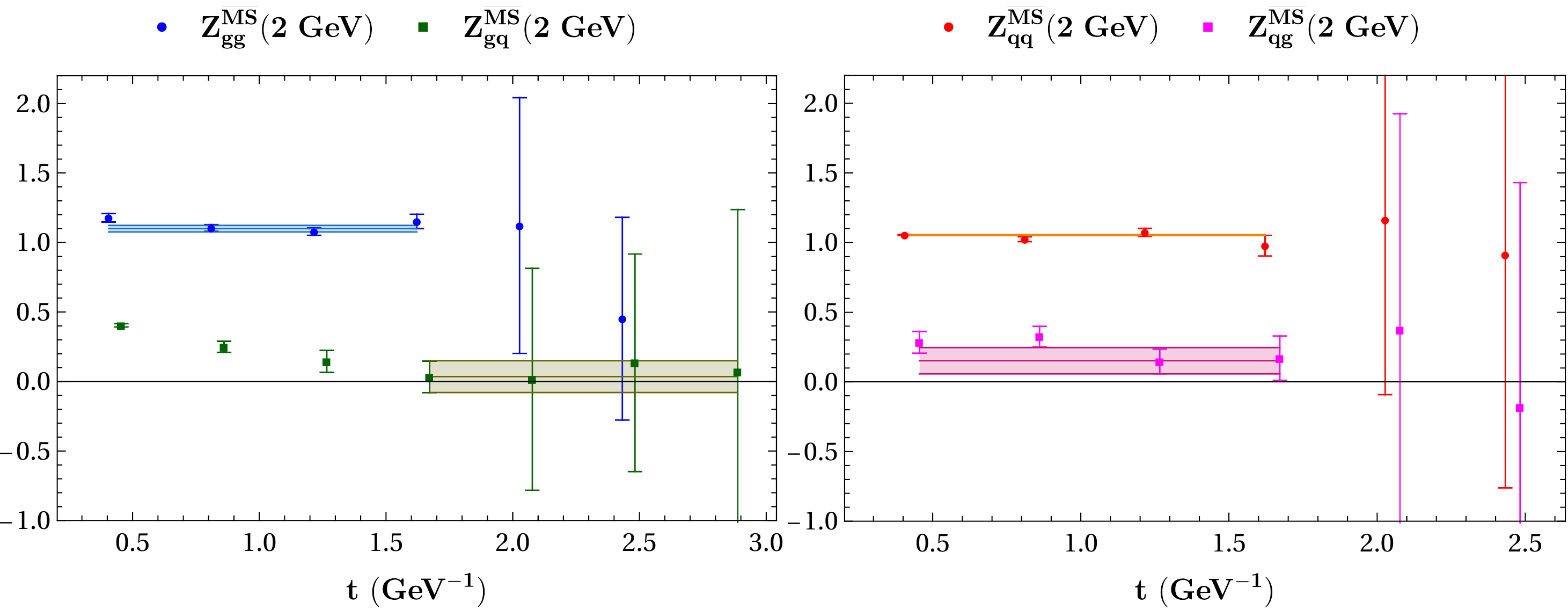}
  \caption{Plots for the mixing coefficients of the non-diagonal ($\mu \neq \nu$) EMT operators in the $\overline{\rm MS}$ scheme for the ensemble cB4.060.12.}
  \label{EMT_mixing_matrix}
\end{figure}

Following a similar procedure as in Sec.~\ref{secII}, we present in Fig.~\ref{EMT_mixing_matrix} preliminary results of all the mixing coefficients in the $\overline{\rm MS}$ scheme for the ensemble cB4.060.12. The final values are listed in Table~\ref{tab:mixing_coeffs}.

\begin{table}
\centering
\begin{tabular} {l | l | l | l | l}
  \hline
  \hline
   Ensemble & $Z_{gg}^{\overline{\rm MS}}$ (2 GeV) & $Z_{gq}^{\overline{\rm MS}}$ (2 GeV) & $Z_{qg}^{\overline{\rm MS}}$ (2 GeV) & $Z_{qq}^{\overline{\rm MS}}$ (2 GeV) \\
\hline
cB4.060.12  & 1.100(23)(125) & 0.035(114)(34) & 0.152(95)(68) & 1.054(4)(5) \\
\hline
\end{tabular}
\caption{Final values of the mixing coefficients of the non-diagonal ($\mu \neq \nu$) EMT operators in the $\overline{\rm MS}$ scheme for the ensemble cB4.060.12.}
\label{tab:mixing_coeffs}
\end{table}

\section{Conclusions and future prospects}
\label{secIV}

In this work, we presented initial results within GIRS for the renormalization of the non-singlet vector one-derivative quark bilinear operator, as well as in the mixing of the traceless EMT operators. We calculated the non-perturbative renormalization functions of these operators in the $\overline{\rm MS}$ scheme. We employed a number of $N_f = 4$  ensembles at 3 different lattice spacings. Our results were applied in the calculation of the quark and gluon average momentum fractions of the nucleon, giving compatible results with previous studies in phenomenology and lattice QCD. 

Some possible extensions/future plans are: \\
$\bullet$ To extract the EMT mixing matrix (using GIRS) for finer lattices (C and D ensembles). \\
$\bullet$ To reduce systematics related to the quark-mass dependence of the RFs. By employing ensembles of the same lattice spacing and different values of the quark mass, we can perform chiral extrapolations in order to obtain RFs in the massless limit. \\
$\bullet$ To extend our calculation to diagonal EMT operators $\mu = \nu$. \\
$\bullet$ To study the renormalization for the trace part of EMT using GIRS, which has some further complications on the lattice.

\acknowledgments \vspace*{-0.2cm}
G.S., J.F., and S.Y. acknowledge financial support from the H2020 project PRACE 6-IP (GA No. 823767) and the EuroCC project (GA No. 951732). H.P. acknowledges support by the European Regional Development Fund and the Republic of Cyprus through the Research and Innovation Foundation (Projects: EXCELLENCE/0918/0066 and EXCELLENCE/0421/0025). The authors gratefully acknowledge computing time allocated on Cyclone supercomputer of the High Performance Computing Facility of The Cyprus Institute under project ID p052, on UCY HPC of the University of Cyprus, as well as computing time granted by the John von Neumann Institute for Computing (NIC) on the supercomputer JUWELS Cluster~\cite{JUWELS_short} at the J\"ulich Supercomputing Centre (JSC).

\bibliographystyle{JHEP}
\bibliography{latt22_GIRS_renormalization.bib}

\end{document}